\documentstyle{article}
\input pz.sty
\input epsf.sty
\begin{document}

\PZhead{3}{30}{2010}{29 April}

\PZtitletl{The light curves of type I\lowercase{a} 
           Supernova 2008\lowercase{gy}}

\PZauth{D. Yu. Tsvetkov$^1$, P. Balanutsa$^1$,
E.S. Gorbovskoy$^1$, D.A. Kuvshinov$^1$, A.V. Krylov$^1$, 
T. Kopytova$^2$, T. Kryachko$^3$, S. Korotkiy$^4$}
\PZinst{Sternberg Astronomical Institute, University Ave., 13,
119992 Moscow, Russia}
\PZinst{Ural State University, Ekaterinburg, Russia}
\PZinst{Astrotel Observatory, Karachay-Cherkessia, Russia}
\PZinst{Ka-Dar Scientific Center and Public Observatory, Moscow, Russia}

\begin{abstract}
CCD $BVRI$ photometry is presented for type Ia
supernova 2008gy. The light curves match the template curves
for fast-declining SN Ia, but the colors appear redder than average,
and the SN may also be slightly subluminous.
SN 2008gy is found to be located far outside the boundaries of three
nearest galaxies, each of them has nearly equal probability to
be the host galaxy. 
\end{abstract}

\begintext

SN 2008gy was discovered by P. Balanutsa on unfiltered CCD images taken 
with the 335-mm telescope near Moscow on October 30.99 UT.
This telescope is a part of the large "MASTER robotic Net" (Lipunov et al.,
2010).   
The new
object was located at 
$\alpha  = 3\hr10\mm00\sec.96, \delta = +19\deg13\arcm23\arcs.1$
(equinox 2000.0), which is $23\arcs$ west and $6\arcs$ north of the 
center of PGC 1584648 (Lipunov, 2008).
Confirmation images were taken by Kryachko and Korotkiy (2008) on
November 2.94 UT with the 80-mm refractor at Karachay-Cherkessia, Russia.
Folatelli et al. (2008) report that they obtained spectra (range 340-602 nm) of 
2008gy with the New Technology Telescope (+EFOSC2) at La Silla on November 
19.0-19.1 UT.
The spectrum of 2008gy shows it to be a type-Ia supernova, two to three
weeks after maximum light, at a redshift of 0.029.

\medskip

SN 2008gy was imaged again with 335-mm telescope near Moscow on
November 7.73 UT. The photometric monitoring of the SN 
in $BVRI$ filters was 
carried out at the 60-cm reflector of Crimean Observatory of
Sternberg Astronomical Institute since November 9 until November 25.
All image reductions and photometry were made using IRAF.\PZfm 
\PZfoot{IRAF is distributed by the National Optical Astronomy Observatory,
which is operated by AURA under cooperative agreement with the
National Science Foundation}

\medskip
\PZfig{12cm}{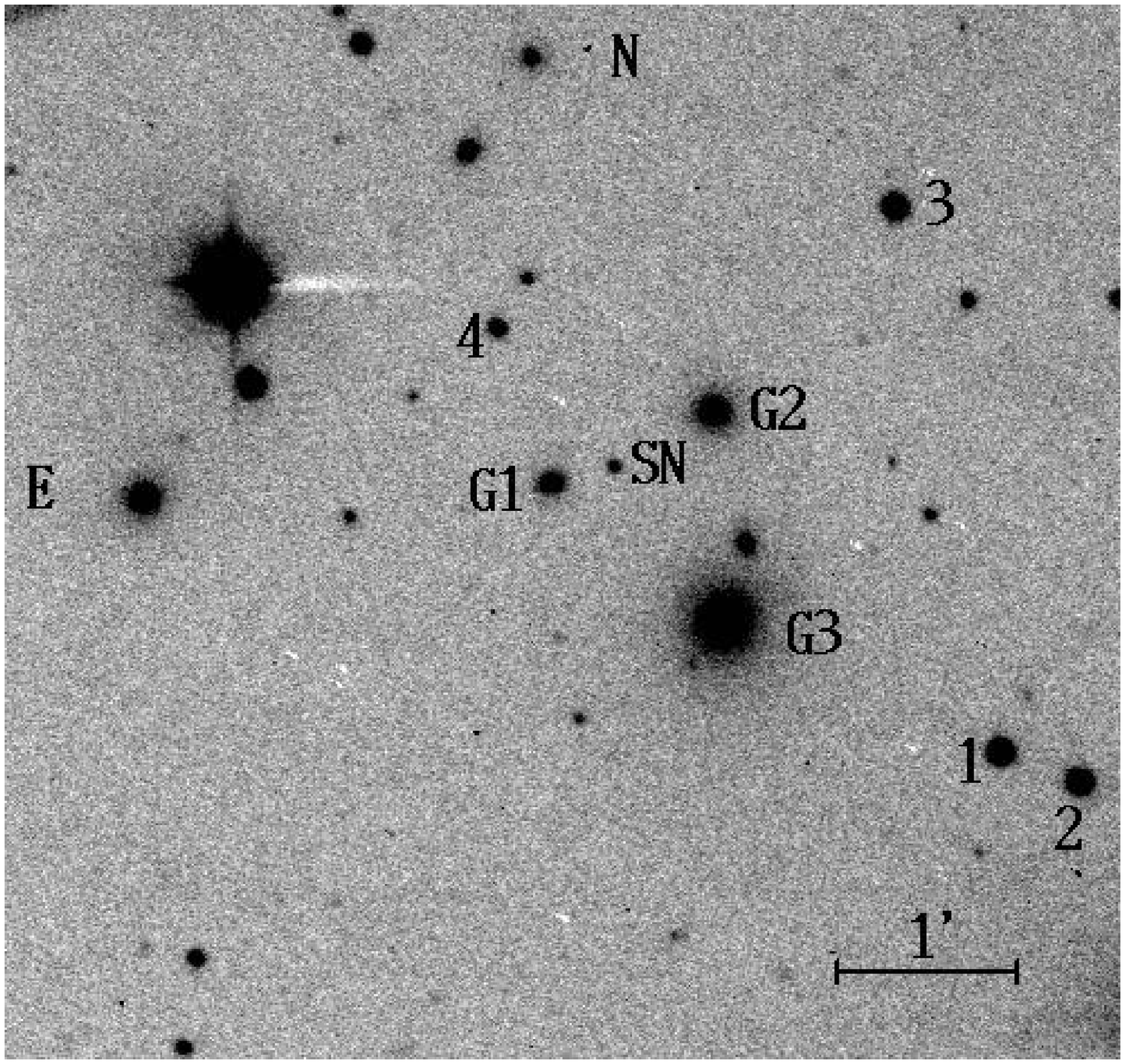}{SN 2008gy, the nearest galaxies and the
local standard stars}

The image of SN 2008gy obtained at the 60-cm reflector in the $R$ band is 
shown in Fig.~1. It is evident that the SN is located in a group
of three galaxies, which are labeled as G1, G2, G3. We searched  
the NED\PZfm\PZfoot{http://nedwww.ipac.caltech.edu}
and Hyperleda\PZfm\PZfoot{http://leda.univ-lyon1.fr} 
databases and found out that G1 is PGC 1584648, G2 is
PGC 1584876, and G3 is IC 1890. Only for IC 1890 the redshift is 
known, which is 0.03412. The value reported
for SN 2008gy differs by 1500 km/s from this estimate,
which seems larger than possible peculiar velocities of galaxies
in a group or the velocity of presupernova inside host galaxy.
But the error of redshift for SN 2008gy is not given, and it is possible
that it can account for most of this difference.
We calculate the angular distances from SN to the three galaxies
on our CCD frames, which are, respectively, for G1, G2, G3: $21\arcs.1$,
$37\arcs.0$, $61\arcs.9$. If we assume that the redshift 0.029 
corresponds to the distance for SN 2008gy and accept the distance
modulus $\mu=35.4$, as reported in NED, the linear projected distances
of the SN from the centers of the galaxies are 12.2 kpc, 21.3 kpc, 35.7 kpc.
According to these data, G1 is the most likely host galaxy. But other 
galaxies are significantly brighter and larger, and the
relative projected radial distances, defined as angular separation divided by
the isophotal radius of the galaxy, for the three galaxies are
2.27, 1.74, 2.06. So, from this point of view G2 is the most likely
host galaxy, and G1 is the least probable host. Taking into account, that
the projected distances are the lower limits to the spatial 
separations, we conclude that there is nearly equal probability for
SN 2008gy to belong to any of the three galaxies, or to be located
in intergalactic medium.      

\medskip

The local standard stars are also marked on Fig.~1. The magnitudes
of these stars are reported in Table~1, they were calibrated on four
photometric nights in November.
It is clear that galaxy background
has no effect on the photometry of SN 2008gy. The magnitudes of
the SN were derived by PSF fitting relative to a sequence of local standard
stars. We used the $R$ magnitudes to calibrate the unfiltered CCD frames
obtained at 335-mm and 80-mm telescopes. 

\begin{table}
\caption{Magnitudes of local standard stars}\vskip2mm
\centering
\begin{tabular}{ccccccccc}
\hline
Star & $B$ & $\sigma_B$ & $V$ & $\sigma_V$ & $R$ &
$\sigma_R$ & $I$ & $\sigma_I$ \\
\hline
1& 15.46& 0.01& 14.71& 0.01& 14.30& 0.01& 13.94& 0.01 \\ 
2& 15.31& 0.02& 14.72& 0.01& 14.34& 0.02& 14.00& 0.01 \\
3& 15.15& 0.02& 14.50& 0.01& 14.11& 0.01& 13.72& 0.02 \\
4& 18.38& 0.26& 17.01& 0.02& 16.19& 0.03& 15.38& 0.03 \\
\hline
\end{tabular}
\end{table}

The results are presented in Table~2 and the light curves are
shown in Fig.~2. 

\begin{table}
\caption{Photometric observations of SN 2008gy}\vskip2mm
\begin{tabular}{cccccccccl}
\hline
JD 2454000+  & $B$ & $\sigma_B$ & $V$ & $\sigma_V$ &  
 $R$ & $\sigma_R$ & $I$ & $\sigma_I$ & Tel. \\
\hline
770.49&       &     &       &     &  17.11& 0.13&       &     &    335-mm \\
773.44&       &     &       &     &  16.96& 0.04&       &     &    80-mm  \\
778.23&       &     &       &     &  17.27& 0.08&       &     &    335-mm \\
780.46&  17.83& 0.07&  17.36& 0.05&  17.26& 0.03&       &     &    600-mm \\
781.42&  18.18& 0.05&  17.58& 0.03&  17.41& 0.03&       &     &    600-mm \\
782.39&  18.17& 0.11&  17.56& 0.05&  17.43& 0.05&       &     &    600-mm \\
786.33&  18.62& 0.06&  17.84& 0.03&  17.75& 0.05&  17.44& 0.17&    600-mm \\
789.38&  19.20& 0.09&  18.02& 0.02&  17.86& 0.03&       &     &    600-mm \\
795.26&  19.81& 0.08&  18.36& 0.02&  17.92& 0.03&  17.51& 0.10&    600-mm \\
796.20&       &     &  18.42& 0.04&  17.79& 0.05&       &     &    600-mm \\
\hline
\end{tabular}
\end{table}

\PZfig{12cm}{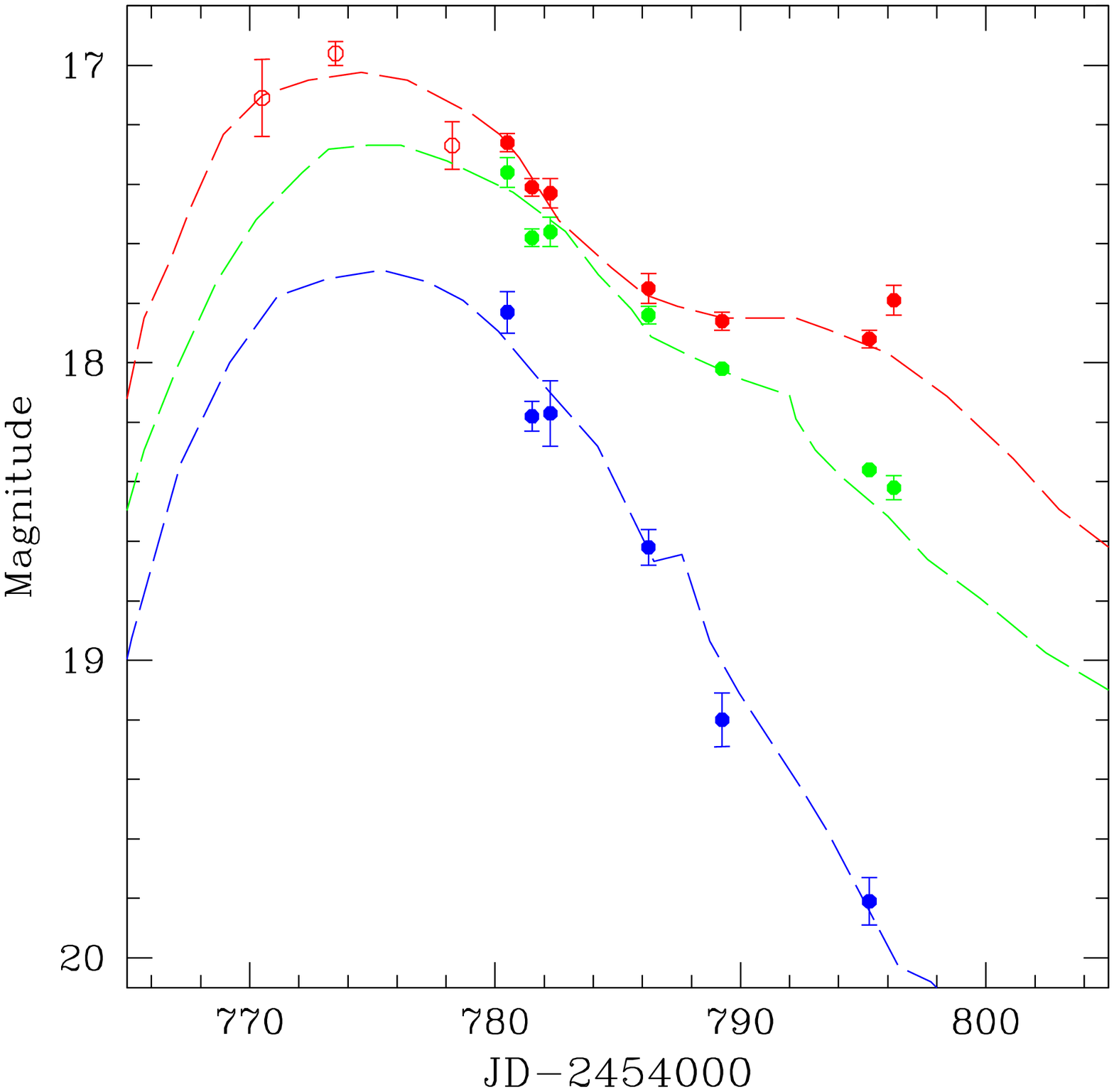}{The light curves of SN 2008gy in the $B$ 
(blue), $V$ (green) and $R$ (red) bands. Dots show data obtained
at the 60-cm telescope, circles are for the unfiltered frames
exposed at the 355-mm and 80-mm telescopes. The dashed lines
are the light curves of SN 1994D}

\PZfig{12cm}{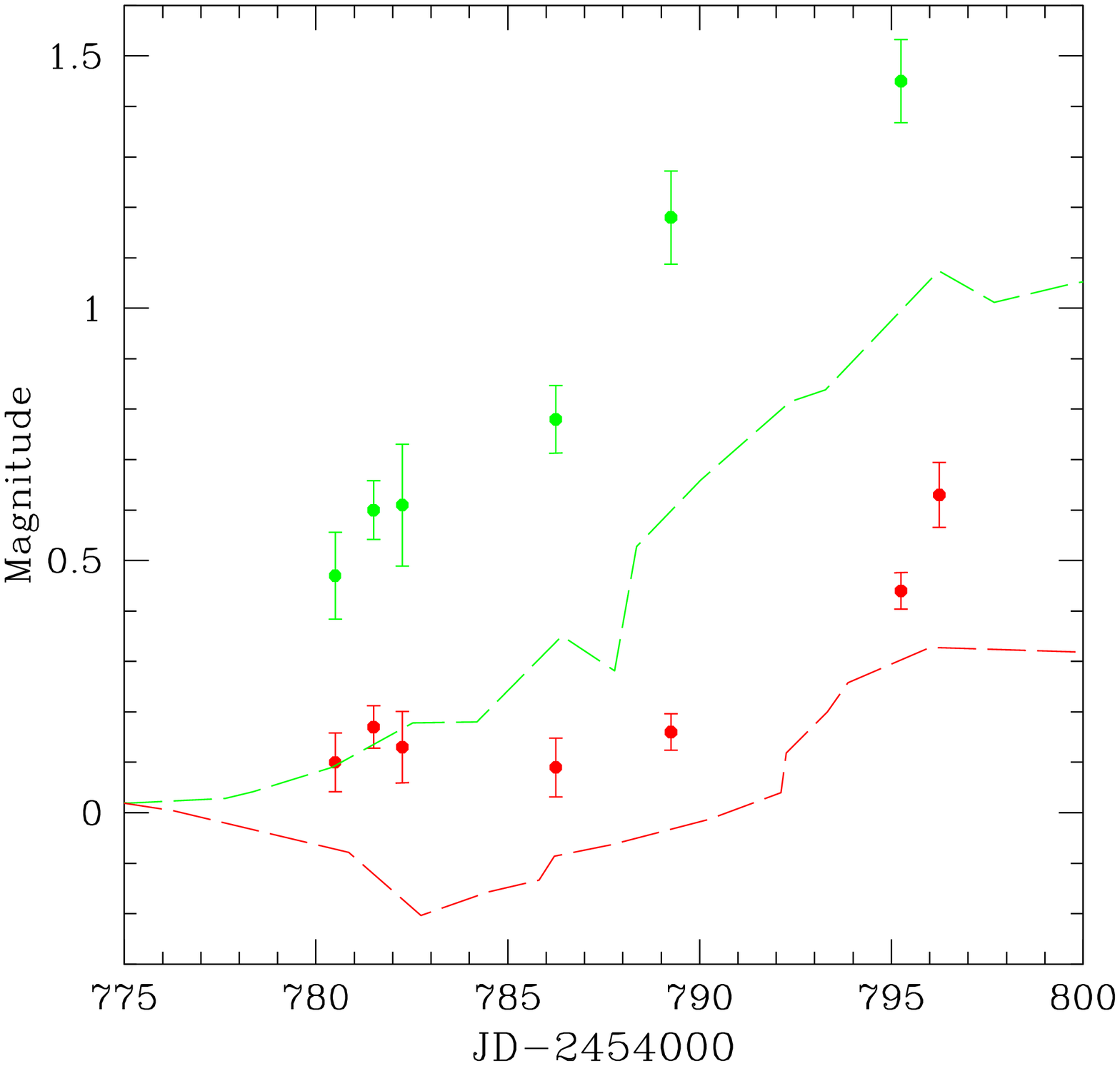}{The $(B-V)$ (green) and $(V-R)$ (red)
color curves for SN 2008gy. The dashed lines are the 
color curves of SN 1994D}

\medskip

The data are best fitted by the light curves of moderately fast-declining 
SN Ia 1994D (Richmond et al., 1995). From this fit we estimate that maximum
light was reached around JD 2454775 (November 4) with $B_{max}=17.7$,
$V_{max}=17.3$, $R_{max}=17.0$. The color curves are shown in Fig.~3.
The colors of SN 2008gy are significantly redder than for SN 1994D,
approximately by 0.44 mag in $(B-V)$ and 0.22 mag in $(V-R)$.
The galactic extinction for SN 2008gy is $E(B-V)_{gal}=0.16$ mag,
$A_B^{gal}=0.7, A_V^{gal}=0.54$, according to NED. We cannon expect
any significant extinction in the parent galaxy, because the SN is 
located outside the borders of all possible host galaxies. 
So, we suppose that SN 2008gy is intrinsically redder than SN 1994D
at the phase of early decline
by about 0.28 mag in $(B-V)$ and 0.11 mag in $(V-R)$. 
Assuming $\mu=35.4$ and correcting only for the galactic extinction,
the absolute magnitudes of SN 2008gy are $M_B=-18.4, M_V=-18.6$.
We may suggest that the rate of early decline for this SN is not
much different from the one for SN 1994D, $\Delta m_{15}(B)\approx 1.35$,
and then the luminosity of SN 2008gy at maximum is significantly 
fainter than can be expected from the relations of 
$M_B$ and $M_V$ versus $\Delta m_{15}(B)$ as presented by 
Hicken et al. (2009). However, the distance modulus corresponding
to the redshift of IC 1890 is 0.35 mag larger, and with absolute
magnitudes brighter by 0.35 mag SN 2008gy fits quite well to the
relation between maximum luminosity and rate of decline.
So, we cannot make definite conclusion on the luminosity of 
SN 2008gy, but its intrinsic red color is much more probable.

\medskip

The possible lower luminosity and red color may be connected with
the location of SN far away from the host galaxy, where the characteristics
of stellar population may be different from those closer to the galaxy
center (see e.g. Sullivan et al. 2010 and references
therein).

\medskip
{\bf Acknowledgements.} 
This work was supported by the Ministry of Science of the Russian Federation 
(state contract No. 02.740.11.0249).
The work of D.T. was partly supported by the 
Leading Scientific Schools Foundation
under grant NSh.433.2008.2 and by RFBR grant 10-02-00249a.
This research has made use of the NASA/IPAC Extragalactic Database (NED) 
which is operated by the Jet Propulsion Laboratory, 
California Institute of Technology, under contract with NASA.
We acknowledge the usage of the HyperLeda database.

\references

Folatelli, G., Forster, F., Stritzinger, M., 2008, {\it CBET}, No. 1577 

Hicken, M., Challis, P., Jha, S., et al., 2009, {\it Astrophys. J.},
{\bf 700}, 331 

Lipunov, V., Kornilov, V., Gorbovskoy, E., et al., 2010, 
{\it Advances in Astronomy}, {\bf 2010}, 1

Lipunov, V.M., 2008, {\it CBET}, No. 1565

Kryachko, T., Korotkiy, S., 2008, {\it CBET}, No. 1565

Richmond, M.W., Treffers, R.R., Filippenko, A.V., et al., 1995,
{\it Astron. J.}, {\bf 109}, 2121 

Sullivan, M., Conley, A., Howell, D.A., et al., 2010, preprint,
arXiv:1003.5119

\endreferences
\end{document}